\documentclass{article}

\PassOptionsToPackage{numbers, square}{natbib}
\usepackage[preprint]{neurips_2022}
\usepackage{neurips_2022}
\usepackage{bmpsize}
\usepackage{microtype}
\usepackage[dvipdfmx]{graphicx}
\usepackage{subfigure}
\usepackage{booktabs} 
\usepackage{adjustbox}
\usepackage{xspace}
\newcommand{\model}{OCTAL\xspace}
\newcommand{\RN}[1]{%
  \textup{\uppercase\expandafter{\romannumeral#1}}%
}
\usepackage{multirow, makecell}
\usepackage{tikz}
\usepackage{tikzscale}
\usepackage[compat=1.1.0]{tikz-feynman}
\usetikzlibrary{shapes,decorations.pathmorphing}
\tikzset{node distance=0.5cm, every edge/.style={draw,
->,thick},snake it/.style={decorate, decoration=snake}}
\tikzstyle{player1}=[circle,draw=black!100,fill=white!20,thick,inner sep=0pt,minimum size=8mm]
\tikzstyle{player3}=[circle,draw=black!100,fill=none,thick,inner sep=0pt,minimum size=10mm]
\tikzstyle{safeReach}=[circle,draw=white!100,fill=white!20,thick,inner sep=0pt,minimum size=8mm]
\tikzstyle{player2}=[rectangle,draw=black!100,fill=white!20,thick,inner sep=0pt,minimum size=7mm]
\tikzstyle{otherLabel}=[circle,draw=black!100,fill=white!20,thick,inner sep=0pt,minimum size=20mm]
\tikzstyle{cpreLabel}=[rectangle,draw=black!100,fill=white!20,thick,inner sep=0pt,minimum size=40mm]

\usepackage[utf8]{inputenc} 
\usepackage[T1]{fontenc}    
\usepackage{hyperref}       
\usepackage{url}            
\usepackage{booktabs}       
\usepackage{amsfonts}       
\usepackage{nicefrac}       
\usepackage{microtype}      
\usepackage{xcolor}         
\usepackage{subfiles}

\usepackage{amsmath}
\usepackage{amssymb}
\usepackage{mathtools}
\usepackage{amsthm}

\usepackage{multirow}
\newcommand{\proj}{OCTAL }

\title{OCTAL: Graph Representation Learning for \\ LTL Model Checking}

\author{
  Prasita Mukherjee, Haoteng Yin, Susheel Suresh, Tiark Rompf \\
  Department of Computer Science, Purdue University\\
  \texttt{\{mukher39,yinht,suresh43,tiark\}@purdue.edu} \\
}

\begin{document}
\maketitle

\begin{abstract}
Model Checking is widely applied in verifying the correctness of complex and concurrent systems against a specification. Pure symbolic approaches while popular, still suffer from the state space explosion problem that makes them impractical for large scale systems and/or specifications. In this paper, we propose to use graph representation learning (GRL) for solving linear temporal logic (LTL) model checking, where the system and the specification are expressed by a B{\"u}chi automaton and an LTL formula respectively. A novel GRL-based framework \model, is designed to learn the representation of the graph-structured system and specification, which reduces the model checking problem to binary classification in the latent space. The empirical experiments show that \model achieves comparable accuracy against canonical SOTA model checkers on three different datasets, with up to $5\times$ overall speedup and above $63\times$ for satisfiability checking alone.
\end{abstract}

\section{Introduction \label{sec:Introduction}}
Model checking, proposed by \citet{clarke2018model} is defined as the problem of deciding whether a specification holds for all executions of a system, where a property is typically specified in a temporal logic like Linear Temporal Logic (LTL), Computational Tree Logic (CTL), etc. Model checking has proven to be extremely effective in verifying complex systems against a set of specifications. It has gained wide acceptance in the fields of systems engineering \cite{fastLTL}, protocol verification in hardware and software systems \cite{kloos2012practical}, malware detection and security protocols \cite{LTLSecurity}. 

Generally, formal specifications are expressed using temporal logic formulae like LTL, CTL, etc. The system/model is expressed using automata like B{\"u}chi \cite{Bchi1990}, Muller, Kripke structures \cite{kripkeStructure}, or, Petri nets \cite{petriNet} to express concurrent systems. Given the system and the specification expressed using temporal logic, model checking can automatically verify whether the system satisfies the specification. The result of model checking on a system expressed in B{\"u}chi automaton will be `1/true' if the automaton satisfies the given specification expressed in temporal logic, and `0/false' otherwise. 

Linear Temporal Logic (LTL) \cite{LTLPnueli} and Computational Tree Logic (CTL) \cite{CTLone,CTLTwo} are the two most widely used temporal logics to express a specification both in industry and academia. Traditionally, LTL model checking first computes the automaton $B_{\neg}$ corresponding to the negation of the given LTL formula, followed by cross product of the system $B$ with the negation $B_{\neg}$, and then checks for the emptiness of the product (illustrated in Figure \ref{fig:prod}). However, this approach suffers from the state space explosion problem \cite{ValmariExplosion} which severely hinders the performance of a model checker, especially when LTL formulae map to exponentially sized systems and the subsequent graph product computation. Methods such as partial order reduction \cite{partialOrderMC}, symmetry \cite{symmetryLTL,symmetryRedLTL}, bounded model checking \cite{Biere99symbolicmodel,OntheFlySymmetryMC}, and equivalence relations have been proposed to address this problem. Some progress has been made by these methods, but the problem of exploded state space remains hard in general. Hence, it still constitutes a major bottleneck in many applications of model checking.

Recently, machine learning (ML) techniques \cite{BehjatiRI,ZhuMC} have gained great success in dealing with the state space explosion problem. \citet{ZhuMC} used machine learning algorithms like random forest, decision trees to predict whether a system (Kripke structure) satisfies a specification (LTL formula). \citet{BehjatiRI} proposed a reinforcement learning based approach for on-the-fly model checking of properties expressed with LTL formulae. Although not guaranteed to be 100\% accurate, leaning-based approaches are much \emph{faster} and \emph{cheaper} compared to traditional model checkers (MCs) as they avoid computing the cross product of the system and specification. Furthermore, it is especially useful in scenarios where traditional MCs time out/fail to solve, while \emph{speed} and \emph{efficiency} are the key factors. For instance, in software verification, traditional MCs are not always a viable choice due to their high cost. Consider a large software development effort, where it would be favorable to use a MC to ascertain a higher degree of correctness than pure testing. In such large-scale deployments, classical MCs often take prohibitively long time for verification, especially when the system under test has an exponential state space. In this case, only ML-based MCs can provide a practical solution, which broadens the applicability of MCs by trading off some amount of accuracy guarantees for better running time and scalability, and is particularly promising for large systems and/or specifications. 

However, previous work of ML-based MCs either does not generalize well across systems and specifications with varying lengths \cite{ZhuMC}, or cater to bounded model checking \cite{BehjatiRI}. Such limitations can be well addressed through more powerful and expressive frameworks, e.g. representation learning. Due to the structural essence of the input, model checking can be naturally formulated into graph tasks. This motivates us to propose a novel graph representation learning (GRL) based framework, \model, to tackle this challenging problem and perform LTL model checking. For a given input, the system is expressed as a B{\"u}chi automaton $B$ and the specification with an LTL formula $\phi$. Then, \model determines whether $B$ satisfies $\phi$ by reducing the problem to binary classification on graph-structured data representing the given system and the specification. Here, the B{\"u}chi automaton is encoded as a bipartite graph and the LTL formula is presented as an expression tree. Instead of modeling them separately, we employ the union operation to obtain their unified graph structures and learn the \textit{joint graph representation} for classification with the output `0/1'. LTL's expression tree representation and graph union operations avoid the construction of system corresponding to the specification, followed by the cross product. When it comes to polynomial sized systems and LTL formulae that map to exponentially sized systems, through this formulation, we can reduce the model checking problem to graph classification that prevents us from encountering the state space explosion problem.

We performed extensive experiments on \model, traditional MCs and neural network baselines for LTL model checking, in terms of both accuracy and speed on three datasets: one from open competition RERS19 \cite{rers2019} and two new ones specifically constructed for this task. Experimental results show that \model consistently achieves $\sim$90\% precision and recall on all three datasets, which indicates its great generalization ability on unseen data and its high utility in practice. Meanwhile, the performance of \model is not negatively affected by the varying length of LTL formulae and systems present in our specially designed dataset \texttt{Diverse} as other ML-based MCs in \cite{ZhuMC} would be. In general, \model is up to $5\times$ faster than the state-of-the-art traditional MCs with over 90\% accuracy, and achieves over $63\times$ speedup in term of inference alone. Our major contribution can be summarized as follows: 1) LTL model checking is firstly formulated as a representation learning task, where the system and the specification are expressed in graph-structured data. 2) A new paradigm of model checking is proposed. To represent and bridge structural/semantic correspondence between LTL and B{\"u}chi automaton, the cross product is replaced by computationally cheaper graph union operations. 3) Two new datasets are constructed for LTL model checking benchmark: \texttt{Short} with large amount of short-size LTL samples and \texttt{Diverse} with more complex and varied-length samples.
\vspace{-3mm}
\section{Preliminaries \label{sec:Prelim}}
\subsection{B{\"u}chi Automata}
In automata theory, a B{\"u}chi automaton (BA) is a system that either accepts or rejects inputs of infinite length. The automaton is represented by a set of states (one initial, some final, and others neither initial nor final), a transition relation, which determines which state should the present state move to, depending on the alphabets that hold in the transition. The system accepts an input if and only if it visits at least one accepting state infinitely often for the input. A B{\"u}chi automaton can be deterministic or non-deterministic. We deal with non-deterministic B{\"u}chi automata systems in this paper, as they are strictly more powerful than deterministic B{\"u}chi automata systems.

A non-deterministic B{\"u}chi automaton $B$ is formally defined as the tuple $(Q, \sum, \Delta, q_{0}, \mathbf{q}_{f})$, where $Q$ is the set of all states of $B$, and is finite; $\sum$ is the finite set of alphabets; $\Delta: Q \times 2^{\sum} \rightarrow 2^Q$ is the transition relation that can map a state to a set of states on the same input set; $q_{0} \in Q$ represents the initial state; $\mathbf{q}_{f} \subset Q$ is the set of final states.

\subsection{Linear Temporal Logic}
Linear Temporal Logic (LTL), is a type of temporal logic that models properties with respect to time. An LTL formula is constructed from a finite set of atomic propositions, logical operators not (!), and (\&), and or ($|$), true (1), false ($N$), and temporal operators. Additional logical operators such as implies, equivalence, etc. that can be replaced by the combination of basic logical operators $(!, \&, |)$. For example, $a \rightarrow b$ is expressed as $!a \: | \: b$.

\begin{figure*}
\centering
\hfill
\subfigure[$B$ accepting `$a \: U \: b$'.\label{FigurelabelBASpec}]{
\begin{minipage}{0.3\textwidth}
\centering    
\includegraphics[height=3.5cm]{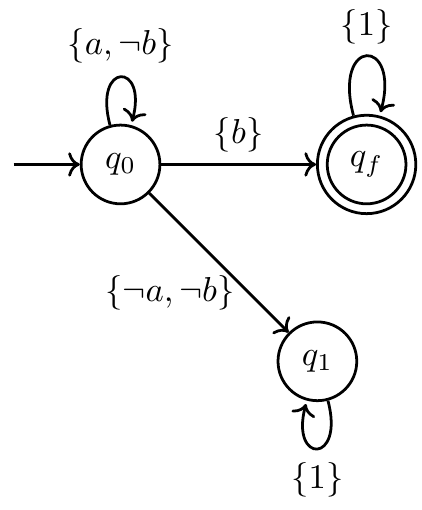}
\vspace{5mm}
\end{minipage}
}
\hfill
\subfigure[$B_{\neg \phi}$ accepting `$!(a \: U \: b)$' \label{FigurelabelNegBASpec}]{
\begin{minipage}{0.32\textwidth}
\centering
\includegraphics[height=3.5cm]{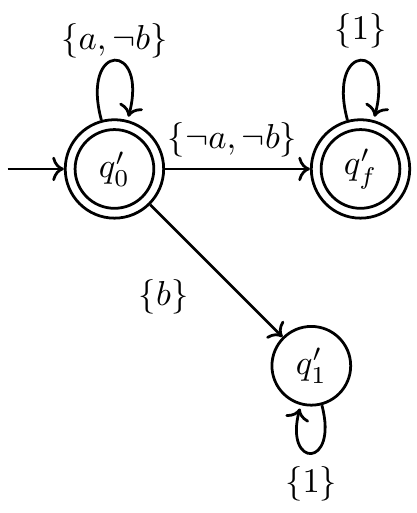}
\vspace{5mm}
\end{minipage}
}
\hfill
\subfigure[$B' = B \times B_{\neg \phi}$\label{FigurelabelBACrossProd}]{
\begin{minipage}{0.33\textwidth}
\centering
\includegraphics[height=4cm]{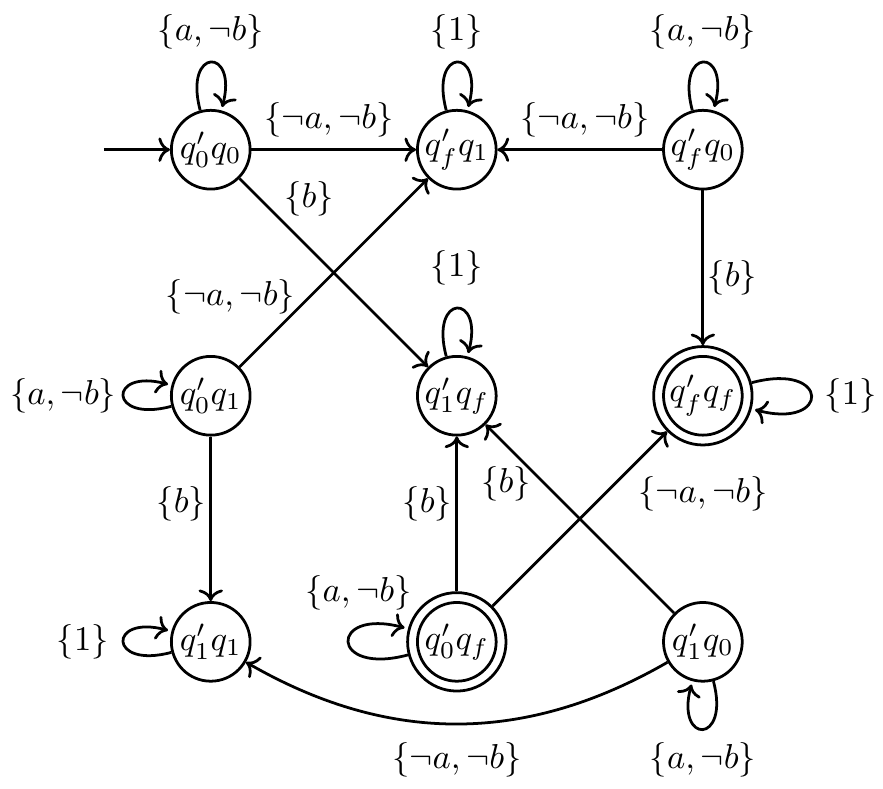}
\end{minipage}
}
\vspace{-3mm}
\caption{Illustration of systems $B$, $B_{\neg \phi}$ and their product $B'$. $B$ accepts `$a \: U \: b$', meaning $a$ must be true until $b$ becomes true. $B_{\neg \phi}$ accepts `$!(a \: U \: b)$', where $\neg \phi$ is the negation of `$a \: U \: b$'. $B'$ is the cross product of $B$ and $B_{\neg \phi}$, which significantly increases the state size. \label{fig:prod}}
\vspace{-3mm}
\end{figure*}

\subsection{LTL Model Checking}
Given a B{\"u}chi automaton $B$ (system), and an LTL formula $\phi$ (specification), the model checking problem decides whether $B$ satisfies  $\phi$. Traditionally, the problem is solved by computing the B{\"u}chi automaton for the negation of the specification $\phi$ as $B_{\neg \phi}$, followed by the product automaton $B' = B \times B_{\neg \phi}$. The problem then reduces to checking the language emptiness of $B'$. The language accepted by $B'$ is said to be empty if and only if $B'$ rejects all inputs. Construction of $B$ is linear in the size of its state space, while $B_{\neg \phi}$ is exponential in the size of $\neg \phi$. The product construction would also lead to an automaton of size $|B| \times 2^{|\neg \phi|}$, which can blow up even $\phi$ is linear in the size of $B$, leading to the state space explosion problem. Figure \ref{FigurelabelBASpec} represents the system $B$ with $\phi$: `$a \: U \: b$', and Figure \ref{FigurelabelNegBASpec} represents $B_{\neg \phi}$. The cross product $B'=B\times B_{\neg \phi}$ is given in Figure \ref{FigurelabelBACrossProd}. $B'$ does not accept anything as there is no feasible path from the initial state ($q'_{0}q_{0}$) to either of the final states ($q'_{0}q_{f}$, $q'_{f}q_{f}$). Here, both $B$ and $B_{\neg \phi}$ are three times smaller than $B'$ in terms of the number of states, and 6 times smaller regarding the number of transitions. As observed in Figure \ref{fig:prod}, it can be concluded that even for moderately complex specifications, the product can still result in an exponential state space, which would severely hinder the performance of traditional model checkers.

\subsection{Graph Neural Networks}
Graph Neural Networks (GNN) extend regular neural networks to better support representation learning in graph domains. A GNN takes the input formed as a graph $\mathcal{G} = (\mathcal{V},\mathcal{E})$, where $\mathcal{V}=[n]$ represents the node set and $\mathcal{E} \subseteq \mathcal{V} \times \mathcal{V}$ is the edge set. The graph can be directed or undirected. Each node $v \in \mathcal{V}$ can be associated with node attributes as $X_v \in \mathbb{R}^{d_v}$. Similarly, each edge $e_{ij}:(v_i, v_j) \in \mathcal{E}$ can also have attributes noted as $E_{ij} \in \mathbb{R}^{d_e}$. Here, $d_v$ and $d_e$ are dimension of node features and edge features, respectively. 

GNNs aim to learn a vector representation $\mathbf{h}_v$ for each node $v$ by aggregating the neighbourhood information and iteratively updating its own hidden representation through $k$ hops as
\[\mathbf{h}_v^k=\text{UPDATE}\left(\mathbf{h}_v^{k-1}, \text{AGGREGATE}\left(\{\mathbf{h}_u^{k-1}|u\in \mathcal{N}_v\}\right)\right).\] 
Here, UPDATE is implemented by neural networks while AGGREGATE is a pooling operation that is invariant to the permutation of the neighbors $\mathcal{N}_v$ regarding node $v$. GNNs combine structural information and node features via message passing on the given graph $\mathcal{G}$. In contrast to standard neural networks, it has the capacity to retain the node representation captured by propagation up to any depth. Recently, GNNs have achieved groundbreaking success on tasks with graph-structured data, especially in node classification \cite{kipf2016semi,hamilton2017inductive,velivckovic2017graph} and graph classification \cite{xu2018powerful,GNNGraphClass}. This motivates us to exploit the structural property of $B$ and  $\phi$, and then represent them as graphs for classification, where the goal is to determine whether $B$ satisfies $\phi$ or not.
\vspace{-3mm}
\section{\model: \texorpdfstring{LTL M\underline{o}del \underline{C}hecking via Graph Represen\underline{ta}tion \underline{L}earning} \label{sec:Method}}

\model determines whether a system, represented by a B{\"u}chi automaton $B$ satisfies a specification $\phi$ represented by an LTL formula through their unified graph representations. We formulate the problem as supervised learning on graphs, where the input $B$ and $\phi$ along with the corresponding label (`0/1') are provided during training, where `1' indicates that $B$ satisfies $\phi$ and `0' otherwise. 


\begin{table}
\centering
\caption{Operands/Variables in specification $\phi$ and system $B$.}
\label{tab:opLogic}
\centering
\resizebox{0.9\columnwidth}{!}{
  \begin{tabular}{l|cccc|ccc}
    \toprule
\multirowcell{2}[-5pt][l]{\textbf{Operands/}\\\textbf{Variables}} & \multicolumn{4}{c|}{Specification $\phi$} & \multicolumn{3}{c}{System $B$} \\ 
\cmidrule(r{0.5em}){2-8}
& \textbf{$\mathcal{A}$} & \textbf{$\mathcal{\tilde A}$} & \textbf{true($1$)} &  \textbf{false($N$)} & \textbf{$\mathcal{A}$} & \textbf{$\mathcal{\neg A}$} & \textbf{true($1$)} \\
\midrule
\textbf{Meaning} & $a$ to $z$ & $!a$ to $!z$ & $a \: | \: !a$, $a \in \mathcal{A}$ & $a \: \& \: !a$, $a \in \mathcal{A}$ & $a$ to $z$ & $\neg a$ to $\neg z$ & $a \: | \: \neg a$, $a \in \mathcal{A}$ \\
\midrule
\textbf{Cardinality} & 26 & 26 & 1 & 1 & 26 & 26 & 1\\
\bottomrule 
\end{tabular}}
\vspace{-3mm}
\end{table}

\subsection{Variables and Operators}
The systems and specifications we deal with are constructed from the operands/variables $\mathcal{A}$, operators $\mathcal{O} =\{G, F, R, W, M, X, U, !, \&, |\}$ and special variables true($1$) and false($N$), the specifics of which are described in Table \ref{tab:opLogic} above and Table \ref{tab:symbol} in Appendix \ref{appx:symb}. Each variable and operator has a distinct meaning and share across $B$ and $\phi$. A variable has its true or negated form (noted as $\mathcal{\tilde A}$ or $\mathcal{\neg A}$).

\subsection{Representation of System and Specification}

\paragraph{System Graph $\mathcal{G}$}
We represent $B$ as a bipartite graph $\mathcal{G} = (V_{\mathcal{G}},E_{\mathcal{G}})$, where $V_{\mathcal{G}} = V_{s} \cup V_{e}$ and $E_{\mathcal{G}} \subseteq V_{s} \times V_{e}$. Here, $V_{s}$ of nodes are the states of $B$, and $V_{e}$ of nodes are the transitions of $B$. There is an edge between $v_i \in V_{s}$ and $v_j \in V_{e}$ if and only if $v_i$ is the source or destination state of the transition $v_j$ in $B$. $\mathcal{G}$ is undirected in nature, but it can be extended to a directed version. 

\begin{figure*}
\centering
\subfigure[$B$ accepting $(a \: U \: !b)$.\label{fig:BA}]{
\begin{minipage}{0.23\textwidth}
\centering    
\includegraphics[height=2cm]{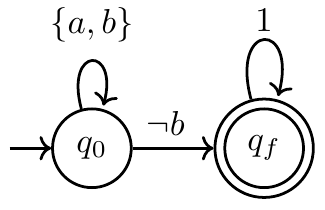}
\end{minipage}
}
\hfill
\subfigure[Bipartite graph $\mathcal{G}$ of $B$.\label{fig:BP}]{
\begin{minipage}{0.23\textwidth}
\centering
\includegraphics[height=2.1cm]{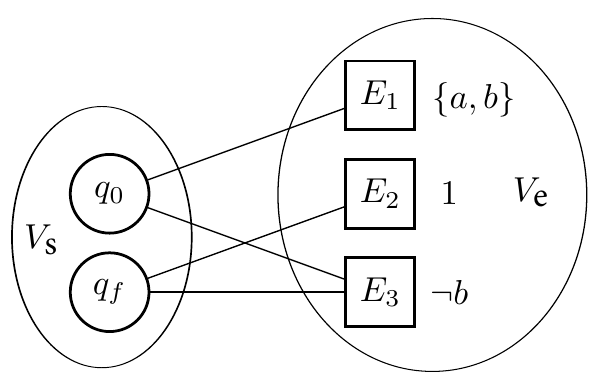}
\end{minipage}
}
\hfill
\subfigure[Expression Tree $\mathcal{T}$ of $\phi: (a \: U \: !b)$.\label{fig:Tree}]{
\begin{minipage}{0.20\textwidth}
\centering
\includegraphics[height=2cm]{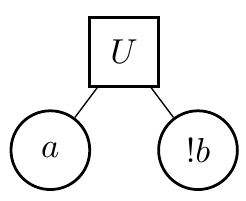}
\end{minipage}
}
\hfill
\subfigure[Unified framework $\mathcal{C}$ as the union of $\mathcal{G}$ and $\mathcal{T}$.\label{fig:Uni}]{
\begin{minipage}{0.25\textwidth}
\centering
\includegraphics[height=2cm]{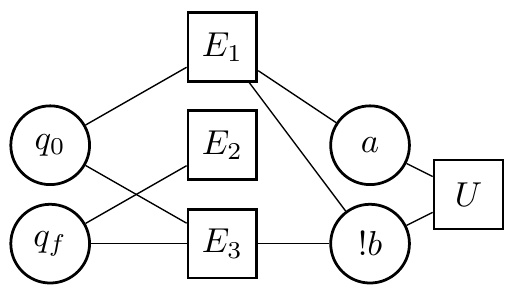}
\end{minipage}
}
\vspace{-3mm}
\caption{Illustration of system $B$, bipartite graph $\mathcal{G}$, expression tree $\mathcal{T}$ for specification $\phi$, and the unified framework $\mathcal{C}$ approximating  $B' = B \times B_{\neg \phi}$, but with polynomial complexity of $|B+\phi|$.}
\vspace{-3mm}
\end{figure*}

Figures \ref{fig:BA} and \ref{fig:BP} illustrate $B$ and the corresponding bipartite graph $\mathcal{G}$. The two states $q_{0}$ and $q_{f}$ form the set $V_{s}$, and the transitions $E_{1}, E_{2}$ and $E_{3}$ form the set $V_{e}$. Since $q_{0}$ is a source and destination state for $E_{1}$, and a source state for $E_{3}$, there is an edge between $q_{0}$ and $E_{1}$, and $q_{0}$ and $E_{3}$ respectively. This is analogously followed by the rest of the graph. The intuition behind representing $B$ as a bipartite graph is to capture the transition labels. Since we aim to learn the overall representation of a given system, both states and transitions in $B$ play an essential role here. A state transits into another state if and only if the transition label is satisfied. To learn the semantic meaning through transitions and their corresponding labels, we map transitions as nodes as shown in Figure \ref{fig:BP} accordingly, and therefore can obtain the representation for them, which is a function of the labels pertaining to the transition.

\paragraph{Specification Graph $\mathcal{T}$}
Every LTL formula can be represented as an expression tree $\mathcal{T} = (V_{\mathcal{T}},E_{\mathcal{T}})$ (see Figure \ref{fig:Tree}), which is constructed based on the precedence and associativity of the operators in LTL formulae, described as follows: 1) $\phi$ is converted to its postfix form, which is used to construct the expression tree; 2) The operators exhibit right associativity, where the unary operators $\{!, G, F, X\}$ have the highest priority. 3) The binary temporal operators $\{U, R, W, M \}$ have the second highest priority, and the boolean connectives $\{\&, |\}$ have the lowest priority. 

$V_{\mathcal{T}}$ constitutes the operators and operands of $\phi$, and $E_{\mathcal{T}} \subseteq V_{\mathcal{T}} \times V_{\mathcal{T}}$. $\phi$ is represented in Negation Normal Form (NNF) \cite{nnf}, which would place $!$ \emph{only} before the operands. This allows us to represent $!a$ as a variable in $\tilde{A}$ and eliminate the $!$ operator. For example, the NNF equivalent of $!(a \: U \:b)$ is $!a \: R \: !b$. Here, the $!$ operator is present only before $a$ and $b$ in the NNF equivalent of the formula $!(a \: U \: b)$. Another compelling reason for representing $\phi$ in NNF is that transitions of $B$ comprise of labels in true($1$), $\mathcal{A}$, and $\mathcal{\neg A}$. Representing negation \emph{only} before variables and eliminating the $!$ operator from $\phi$ enables the shared representation for variables across $B$ and $\phi$. As a result, there is no semantic difference between $\mathcal{\neg A}$ and $\mathcal{\tilde A}$: $\neg a \in \mathcal{\neg A}$ may occur in a transition label of $B$ and $!a \in \mathcal{\tilde A}$ may occur in a leaf node of $\phi$, but both $\neg a$ and $!a$ signify that $a$ does not hold.  

\subsection{Bridging System and Specification via Graph Union}
To establish the relation between graphs of system $\mathcal{G}$ and specification $\mathcal{T}$, we propose to construct the unified graph $\mathcal{C}$ to model them jointly. As Figure \ref{fig:prod} shows, traditional approaches of model checking computes the intersection of $B$ and $B_{\neg \phi}$ by their cross product, which results in an automaton $B'$ whose states are the product of the states of $B$ and $B_{\neg \phi}$, and its transitions depend on the transition labels of both $B$ and $B_{\neg \phi}$. Since our main goal is to avoid constructing $B_{\neg \phi}$ and thus $B'$, we directly feed the input graph without products to \proj and aim to use neural networks to learn the latent correspondence by combining $\mathcal{G}$ and $\mathcal{T}$ as a joint framework $\mathcal{C}$ in the following way. Each transition label consists of operands and variables in $\mathcal{A}$ or $\mathcal{\neg A}$, which is shared across the system and specification. Based on this observation, we join graphs $\mathcal{G}$ and $\mathcal{T}$ by adding a link between the corresponding nodes $V_e \in \mathcal{G}$ and $V_\mathcal{T} \in \mathcal{T}$ if they contain the same variable/operand that belongs to $\mathcal{A}$ or $\mathcal{\tilde A}$/$\mathcal{\neg A}$. Figure \ref{fig:Uni} shows the result of such graph union between $\mathcal{G}$ and $\mathcal{T}$. Here, there is an edge between $a$ and $E_{1}$ as $a$ is contained in $E_{1}$. Similarly, there is one edge between $!b$ and $E_{1}$, and the other between $!b$ and $E_{3}$ as $b$ is in $E_{1}$ and $\neg b$ is in $E_{3}$.

Formally, the unified framework is a joint graph $\mathcal{C} = (V_{\mathcal{C}},E_{\mathcal{C}})$ represented as the union of $\mathcal{G}$ and $\mathcal{T}$, where $V_{\mathcal{C}} = V_{\mathcal{G}} \cup V_{\mathcal{T}}$, and $E_{\mathcal{C}} \subseteq V_{\mathcal{C}} \times V_{\mathcal{C}}$ such that, there is an edge between every leaf node $l \in V_{\mathcal{T}}$, and nodes $E \in V_{e}$ such that $l$ contains a variable $a \in \mathcal{A}$ or $!a \in \mathcal{\tilde A}$, and $E$ also contains the same/equivalent variable $a \in \mathcal{A}$ or $\neg a \in \mathcal{\neg A}$.

\subsection{Node Encoding \label{sec:encoding}} 
Each node $v \in \mathcal{C}$ consisting of operands/variables is represented by a vector as follows

\[
      \{\substack{\RN{1} \\ \underbrace{[\_]}_{1/0}}
      \substack{\RN{2} \\ \underbrace{[\_,\_,\_,....,\_]}_{\forall a \in \mathcal{A}}}
      \substack{\RN{3} \\ \underbrace{[\_,\_,\_,....,\_]}_{!a/\neg a, \forall a \in \mathcal{A}}}
      \substack{\RN{4} \\ \underbrace{[\_,\_,\_,....,\_]}_{ \forall o \in \{\mathcal{O} / !\}}}
      \substack{\RN{5} \\ \underbrace{[\_,\_]}_{q \in Q} \}}
\]

Part I of 1 bit is reserved for the special variable \textbf{true}($1$) or \textbf{false}($N$). Part II encodes $\forall a \in \mathcal{A}$, with size of $|\mathcal{A}|$. Part III of size $|\mathcal{A}|$ encodes variables/operands in either $\mathcal{\neg A}$ or $\mathcal{\tilde A}$, as both of them are semantically equivalent. Part IV corresponds to operators in $\mathcal{O}$ except $!$, with the size of $|\mathcal{O}|-1$. The last part V represents the type of state $q$ of $B$ in 2 bits. Part I and V use one-hot encoding for indication. Each variable/operand has a distinct meaning but sharing across $\mathcal{G}$ and $\mathcal{T}$. Without loss of generality, the value of $ i \in \mathcal{A}\cup\mathcal{O}$ is sampled from a normal distribution $\mathcal{N}(\mu_{i},\sigma)$ with a distinct mean $\mu_i$ and the controlled variance $\sigma$ to avoid overlaps. The true form $a$ and negated form $!a / \neg a$ of a variable share the same value but are located in different sections ($\mathcal{A}$ in Part II, $\neg \mathcal{A}$/$\tilde{\mathcal{A}}$ in Part III).

\subsection{Learning Unified Graph for Modeling Checking}
GNNs as a powerful tool can capture both structural information and node features for graph-structured data through propagation and aggregation of information in local neighborhoods. The message passing framework of GNNs is ideal for exploiting the structural correspondence between the automaton and LTL formula as well as the semantic meaning of transitions. Graph Isomorphism Network (GIN, \cite{xu2018powerful}), one of the most expressive GRL models, is employed to learn the representation of the unified framework $\mathcal{C}$. GIN has demonstrated great performance in graph classification tasks and is expected to obtain better representation of graphs with complex structures. The key intuition here is to jointly learn the representation $\mathcal{G}$ and $\mathcal{T}$. Since two structurally similar $B$'s (or $\phi$'s) can represent different behaviors depending on the contents of the transition, both structure and labels that describe the semantics of the system (or specification) are equally important for LTL model checking. Modeling the system or the specification separately would lose such crucial connection between them, which is also the essential component in traditional model checkers formed as graph products. Hence, we deploy GIN to capture both structure and semantics of $B$ and $\phi$ jointly in the representation learnt, and the significance of the joint representation for $\mathcal{C}$ is further solidified in Sections \ref{sec:perf} and \ref{sec:case}.
\vspace{-3mm}

\section{Evaluation \label{sec:exp}}
\subsection{Dataset \label{sec:data}}
We present three datasets for LTL model checking: \texttt{Short} and \texttt{Diverse} are created through Spot \cite{spotpaper}, while \texttt{RERS19} is obtained from the annual competition RERS19 \cite{rers2019}. The data generated from Spot and provided by RERS19 are specifications, i.e., a set of LTL formulae $\phi$'s. To generate the corresponding automaton $B$ for each $\phi$, the tool LTL3BA \cite{ltl3bapaper} is used, due to its superior speed \cite{ltl3baspeed}. Dataset \texttt{Short} consists of short-size specifications, all of which can be solved by LTL3BA within the time limit (2 mins). \texttt{Short} aims to test the checking speed of models w.r.t LTL3BA. Dataset \texttt{Diverse} comprises of complex and lengthy specifications, which LTL3BA times out for certain instances. \texttt{Diverse} is designed to test model's generalization ability for varied length of specifications, along with the speedup. \texttt{RERS19} is a standard benchmark for sequential LTL from the RERS19 competition. The statistics of datasets are summarized in Table \ref{tab:data}, and detailed in Appendix \ref{appx:data}.

\begin{table}[tp]
\centering
\caption{Summary statistics of datasets.}
\label{tab:data}
\centering
\resizebox{0.65\columnwidth}{!}{
  \begin{tabular}{llllc}
    \toprule
    \textbf{Dataset}&\textbf{Len\_LTL} &\textbf{\#State}&\textbf{\#Transition} &\textbf{\#Sample}\\
    \midrule
    \texttt{Short} &  [1 - 80]  & [1 - 95] & [1 - 1,711] & 9,424 \\
    \texttt{Diverse} & [1 - 144] & [1 - 2,234] & [1 - 397,814] & 6,107 \\
    \texttt{RERS19} & [3 - 39] & [1 - 21] & [3 - 157] & 1,798 \\
  \bottomrule 
\end{tabular}}
\vspace{-3mm}
\end{table}

\subsection{Experimental Settings \label{sec:settings}}
\paragraph{Training}
\model is trained on \texttt{Short} and \texttt{Diverse} datasets with an 80-20 split between training and validation sets, which contain \emph{equal} number of positive and negative samples for classification, and are randomly shuffled before splitting. We use Adam \cite{AdamOptimizer} to optimize model parameters with the initial learning rate \texttt{lr=1e-5}, dropout value of 0.1, and the Binary Classification Entropy as the loss function for all our experiments. Early stopping is adopted when the highest accuracy on validation no longer increases for five consecutive checkpoints. All experiments are run 5 times independently, and the average performance and standard deviations are reported. Additional experiments and analysis including generalization of different models on \texttt{Diverse}, one-hot node encoding scheme and the directed version of $\mathcal{G}$ are presented in Appendix \ref{appx:res}.

\paragraph{Evaluation Metrics}
Two settings are designed to evaluate the model: binary classification and one-vs-many. Accuracy, Precision and Recall are used for binary classification of pair-wise LTL-BA.
The ranking metrics Hits@K and Mean Reciprocal Rank (MRR) are adopted for one-vs-many case. Hit@K counts the ratio of positive samples ranked at the top-K place against all the negative ones. MRR firstly calculates the inverse of the ranking of the first correct prediction against the given negative samples (50 by default), and then an average is taken over the total candidates.

\paragraph{Baselines} Two class of methods are selected to compare for LTL model checking:

\textit{Traditional Model Checkers} LTL3BA, LTL2BA \cite{ltl2bapaper}, Spin \cite{spinMC} and Spot are the SOTA tools that perform the traditional way of symbolic model checking. These tools perform a set of optimizations on top of traditional model checking algorithms, and are capable of solving complex systems and specifications of varying lengths. LTL3BA is the fastest tool among them in terms of generating $B$ corresponding to $\phi$ and checking for their equivalence. Therefore, we select LTL3BA as the strongest baseline for speed test: run it with a timeout of 2 minutes per LTL-BA pair, and the pairs it fails to model check within the time constraint are marked with the output `unknown' and recorded the run-time as the upper limit.

\textit{Learning-based Models} All neural networks take the unified framework $\mathcal{C}$ as input, and outputs `0/1':
\textbf{MLP} A multilayer perceptron (MLP, \cite{MLPCite}) classifier directly uses the given raw node features as input without utilizing graph structure of the unified framework $\mathcal{C}$ for model checking.

\textbf{LinkPredictor} Graph Convolutional Network (GCN, \cite{kipf2016semi}) is used to learn representations of $\mathcal{G}$ and $\mathcal{T}$ \emph{separately}. The obtained graph embeddings of $\mathcal{G}$ and $\mathcal{T}$ are multiplied/concatenated and then fed into linear layers for classification, which essentially converts the satisfiability checking problem between $\mathcal{G}$ and $\mathcal{T}$ to a link prediction task.

\textbf{\proj} A GRL framework that reduces the model checking to a graph classification task by \emph{jointly} learning the representation of $\mathcal{G}$ and $\mathcal{T}$ through the unified framework $\mathcal{C}$. The resultant node embeddings are aggregated using mean pooling to obtain the embedding for the unified graph, which is later fed into an MLP classifier for prediction. There are two variants: the default one uses 3-layer GIN noted as \proj(GIN), the other adopts 3-layer GCN as \proj(GCN).

\begin{table}
\caption{Classification accuracy and precision/recall for LTL model checking. \texttt{RERS19(S)} and \texttt{RERS19(D)} represent LTL-BA being trained on \texttt{Short} and \texttt{Diverse}, and then tested on \texttt{RERS19}.}
\label{tab:acc}
\label{tab:precrecall}
\vspace{-2mm}
\centering
\renewcommand{\arraystretch}{1.15} 
\resizebox{0.75\columnwidth}{!}{
\begin{tabular}{ll|cc|cc}
\toprule
    & \textbf{Models} & \texttt{Short} & \texttt{Diverse} & \texttt{RERS19(S)} & \texttt{RERS19(D)} \\
\midrule
    \multirow{3}{*}{\rotatebox{90}{\texttt{Accuracy~}}} & \textbf{MLP} &  65.42±1.75 & 58.13±1.13 & 64.80±4.28 & 48.96±2.37\\
    & \textbf{LinkPredictor} & 89.57±0.76 & 73.73±1.53 & 84.60±1.17 & 75.00±2.32 \\
    & \textbf{OCTAL(GCN)} & 96.02±0.45 & 85.91±1.04 & 91.58±0.63 & 87.02±4.10\\
    & \textbf{OCTAL(GIN)} & \textbf{96.62±0.31} & \textbf{86.74±0.84} & \textbf{94.09±1.34} & \textbf{94.20±0.88}\\
\midrule
\midrule
    \multirow{3}{*}{\rotatebox{90}{~~~~\texttt{OCTAL}}} & \textbf{Precision} &  96.43±1.33 & 84.50±1.15 & 98.07±0.48 & 91.35±1.04\\
    & \textbf{Recall} &  96.86±0.97 & 88.70±0.91 & 89.97±3.00 & 97.64±1.19\\
\bottomrule
\end{tabular}}
\vspace{-2mm}
\end{table}

\begin{table}
\caption{Running time (inference) of different models for LTL model checking.}
\label{tab:complexity}
\vspace{-2mm}
\centering
\resizebox{0.75\columnwidth}{!}{
\begin{tabular}{l|c|ccc|c}
\toprule
    \textbf{Dataset} &  \textbf{LTL3BA} & \textbf{MLP} & \textbf{LinkPredictor} & \textbf{\proj} & \textbf{Overhead}\\
    \midrule
    \texttt{Short} & 91s & \textbf{0.28s} & 0.39s & 0.38s & 33s \\
    \texttt{Diverse} & 43m & \textbf{0.43s} & 0.74s & 0.55s & 9m\\
    \texttt{RERS19} & 17s & 0.29s & \textbf{0.22s} & 0.27s  & 6.3s\\
\bottomrule
\end{tabular}}
\vspace{-3mm}
\end{table}

\subsection{Performance Analysis \label{sec:perf}}
Table \ref{tab:acc} (upper) shows the performance of different methods for LTL model checking as a classification task. \proj(GIN) consistently outperforms all the other baselines, and achieves above 90\% accuracy on three of four scenarios. \proj(GCN) also obtains above 85\% accuracy, but on average slightly behind \proj(GIN), due to GCN's limited expressiveness compared to GIN. In general, message passing frameworks of \proj perform much better than MLP and LinkPredictor, which either do not take graph structures into account or model systems and specifications separately. This validates the importance of \proj that jointly learning the representation of both system and specification, and their semantic correspondence under the unified framework via message passing. 

To better evaluate the generalization capability of models, the domain shift learning between datasets is considered: \texttt{Short}$\to$\texttt{RERS19} as \texttt{RERS19(S)} and \texttt{Diverse}$\to$\texttt{RERS19} as \texttt{RERS19(D)}. Comparing the left and right side results in Table \ref{tab:acc}, there is certain performance decay (5\% $\sim$ 10\% drops) from MLP and LinkPredictor, which demonstrates their poor generalization when the domain shifts, and thus indicates their limited practical value. \model generally maintains or even exceeds performance after domain shifting, and significantly outperform other two baselines (over 7\% gap). This implies great stability and generalization of \model between datasets with different patterns of systems and varying-length of LTL formulae, which is one of key metrics for real applications in production line. To further validate the effectiveness of \model for practical use, we present the metrics precision and recall in Table \ref{tab:precrecall} (down). \model consistently achieves around 90\% precision and recall across all datasets. As the result suggests, \model has a low false positive rate, and can accurately retrieve over 88\% true positive samples from all three test sets, thus illustrating its usefulness in real applications, such as software verification (case study in Section \ref{sec:case}).

\subsection{Runtime and Complexity Analysis \label{sec:complex}}
Table \ref{tab:complexity} shows the runtime of both the tradition and learning-based model checkers for satisfiability checking of all samples in the listed dataset.  
For fair comparison, we include the graph preprocessing time as the overhead for neural network (NN) models since they take the unified framework $\mathcal{C}$ as input, where constructing graphs $\mathcal{G}$ and $\mathcal{T}$ is required in prior. NN-based models show similar runtime as all of them take less than 1s for inference. Compared to traditional model checkers, with preprocessing overhead considered, NN-based models are still much faster than LTL3BA under every scenario. \proj as the most accurate NN models, achieves speedups of $\sim3\times$, $\sim5\times$ and $\sim3\times$ over LTL3BA, on \texttt{Short}, \texttt{Diverse} and \texttt{RERS19}, respectively. It is worth noting that, in terms of inference time alone, \proj is $\sim239\times$, $\sim4690\times$, and $\sim63\times$ faster than LTL3BA on these three datasets, respectively. This leads us to conclude that \proj can outperform the SOTA traditional model checkers with respect to speed, along with consistent accuracy across different datasets. Note that, as a proof of concept, the graph preprocessing time presented above is not extensively optimized in terms of speed. We aim to provide a parallel graph construction algorithm later that would significantly reduce the preprocessing overhead. 

Traditional model checkers need map $\neg \phi$ to $B_{\neg \phi}$, compute the product $B' = B_{\neg \phi} \times B$, and then check emptiness of $B'$. The use of the union operation pairing with GRL framework enables \proj to avoid such non-polynomial complexity of graph products. Accordingly, the complexity of our proposed method is  reduced to polynomial in the size of $|B + \phi|$. In the Sequential LTL Track of RERS19 competition, the top three teams managed to solve 100\%, 65\% and 25\% of the given problems, respectively. The results presented did not include time, hence we are unable to make a comparison with respect to runtime for \proj against the leaderboard. However, we can still conclude that \proj beats the 2nd and 3rd teams by a large margin in terms of the number of problems it correctly solved (>90\%) combining with its superior runtime and efficiency.

\begin{table*}[t]
\caption{LEFT: One-vs-many results of different models for LTL model checking. RIGHT: Classification accuracy on \texttt{Short} with structural perturbation (Edges are dropped under certain probability).} 
\label{tab:ranking}
\centering
\resizebox{0.95\columnwidth}{!}{
\begin{tabular}{l|cc|cc|ccc}
\toprule
   \multirow{2}{*}{\textbf{Models}} & \multicolumn{2}{c}{\texttt{RERS19(S)}}  & \multicolumn{2}{c}{\texttt{RERS19(D)}} & \multicolumn{3}{|c}{\texttt{Short} with Structural Perturbation}\\ \cmidrule(r{0.5em}){2-8}
    & MRR & Hits@3  & MRR & Hits@3 & \texttt{30\%} & \texttt{50\%}& \texttt{100\%}\\
    \midrule
    \textbf{MLP} & 24.15 & 25.58 & 12.02 & 10.17 & 50.00±0.00 & 50.00±0.00 & 50.00±0.00\\
    \textbf{LinkPredictor} & 55.77 & 73.08 & 31.77 & 35.50 & 50.00±0.00 & 50.00±0.00 & 50.00±0.00\\
     \textbf{\proj(GCN)} & 73.41 & 88.64 & 65.14 & 80.47 & 61.75±0.20 & 61.75±3.20 & \textbf{51.93±0.40}\\
    \textbf{\proj(GIN)} & \textbf{95.74} & \textbf{97.78} & \textbf{77.17} & \textbf{88.21} & \textbf{83.59±1.60} & \textbf{79.44±0.23} & 50.52±0.39 \\
\bottomrule
\end{tabular}}
\vspace{-3mm}
\end{table*}

\subsection{Case Study: Ranking and Robustness in Software Verification \label{sec:case}}
Table \ref{tab:ranking} LEFT shows the results for LTL model checking on one-vs-many case. The test set \texttt{RERS19} is constructed such that for every $B$, there is only one $\phi$ where $B$ does satisfy and fifty other $\phi$'s do not. Hence, every system is paired with one positive and fifty negative specifications. The goal of this task is to examine whether a model could distinguish the positive sample out of many changeling negative ones, which is imitating real scenarios in software verification, by which \model is strongly motivated. The metric Hit@3 is used to count the percentage of positive samples ranked top-3 against all the negative ones. We observe a large margin (consistently over 15\%) between two NN-based models and two \proj variants, especially on \texttt{RERS19(D)}. Furthermore, \proj outperforms LinkPredictor and MLP on all metrics. This again demonstrates the inherent limitation of graph-agnostic models that fail to capture the semantic correspondence between the system and the specification. The analysis also leads us to observe the generalization power of \proj, as it can identify a positive sample with very high confidence among a bunch of negative samples, thus making it feasible to be deployed in real world applications of software development.

Table \ref{tab:ranking} RIGHT shows the results when a percentage of edges is randomly dropped from the input graph. This case evaluates the robustness of \proj as well as the significance of the unified framework $\mathcal{C}$. 
The positive and negative pairs of \texttt{Short} are generated as follows: if $B_{r}$ satisfies $\phi_{r}$, then $\langle B_{r}, \phi_{r}\rangle$ is a positive pair. Let $p\%$ of transitions (edges) in $\mathcal{G}$ be dropped from $B_{r}$, and the modified automaton be $B_{r}'$. $B_{r}'$ does not accept $\phi_{r}$ as all constructed automata do not contain redundant transitions. Thus, dropping even one transition would lead $\langle B_{r}', \phi_{r}\rangle$ to be a negative pair. When $p=100\%$ corresponds to the case that all the edges in $\mathcal{G}$ are dropped, then transition nodes $V_{e}$ and state nodes $V_s$ are completely isolated.
Results from Table \ref{tab:ranking} clearly depict the importance of graph structure as MLP and LinkPredictor fail to generalize with perturbed input and give random guess under this setting. Since they completely overlook or only partially consider the structure and connection within the graph input, MLP and LinkPredictor cannot tell the difference between positive and negative samples: the input are identical for graph-agnostic models with or without edges. Therefore, just relying on node features is not a feasible solution when edges are dropped or $\mathcal{G}$ is corrupted this way. On the other hand, message passing frameworks perform better as \proj(GIN) still holds the distinguishing ability for perturbation and achieves good generalization when the dropping range is between $30\%$ and $50\%$. When $p=100\%$, every transition in $B$ is dropped, thus also leading to poor performance, that validates the importance of structural information for model checking. When $p<30\%$, $B_r$ and $B_r'$ have large portion structural overlaps, leading to similar latent representations learnt by GNN models. This setting is very challenging and uncommon in practice, as different software implementations would lead to more than moderate structure changes in the system. Nevertheless, \model can be further enhanced by adding regularization terms or via adversarial training to obtain high-level sensitivity for such special needs, which we leave for future study.
\vspace{-3mm}
\section{Related Work \label{sec:Related}}
To the best of our knowledge, this is the first work that applies graph representation learning to solve LTL-based model checking. Previously, the most relevant work to us was using GNNs to solve SAT for boolean satisfiability \cite{neuroSAT}, satisfiability of 2QBF formulae \cite{graphqbf}, and automated proof search \cite{HOLGraph} in the higher order logic space. Regarding model checking, learning-based approaches have been used to select the most suitable model checker for an appropriate property and program \cite{MUXTulsianKKLN14}. \citet{IntegrateBorgesGL10} attempts to learn how to reshape a system expressed by NuSMV \cite{cimatti1999nusmv} to satisfy the property in temporal logic. \citet{TemporalTransformer} trains a transformer to predict a satisfiable trace for an LTL formula. 
\citet{ZhuMC} proposes model checking based on binary classification of machine learning. They represent the system using Kripke structures and specification using LTL formulae, and then served as input features to supervised learning algorithms, such as Random Forest, Decision Trees, Boosted Tree and Logistic Regression, which achieve similar accuracy as classical LTL model checking counterparts while partially avoiding the state space explosion problem. However, these ML-based algorithms do not generalize well when tested with formulae of varying lengths. \citet{BehjatiRI} proposes a reinforcement learning based approach for on-the-fly LTL model checking, which is designed to look for invalid runs or counterexamples by awarding heuristics with a agent. Their approach performs much faster than the classical model checkers and can verify systems with large state spaces, but the state space that an agent can reach is still bounded.
\vspace{-3mm}
\section{Conclusions and Future Work \label{sec:Conclusion}}
\proj is a novel graph representation learning based framework for LTL model checking. It can be extremely useful for the first line of software development cycle, as it offers reasonable accuracy and robustness for early and quick verification of model checking compared to time-consuming unit tests and other efforts in ensuring the correctness of a given system. \model is not intended to replace traditional model checkers, rather it makes model checking affordable and scalable for scenarios where traditional model checkers are infeasible. It would broaden the scope and use of model checking on a variety of less safety critical systems, by providing low-cost inference with high accuracy and robustness. It can also be enhanced with guarantee provided by applying traditional model checkers to limited candidates filtered by \model.
 
 In future, we would like to extend \proj to study its generalization for semantically equivalent formulae that may or may not vary in size and variables. In addition, it is very promising for \proj to support the generation of a counter example trace for the `no' answers. Since the counter example generation is a relatively easier problem, tentatively, the user can invoke a traditional model checker to obtain it. In fact, they would only pay a low cost by deploying \model as the ratio of false negatives for the domain shifting task \texttt{RERS19(S)} and \texttt{RERS19(D)} is less than 10\% and 3\%, respectively. Meanwhile, recent developments such as GNNExplainer \cite{GNNExplainer} and Gem \cite{lin2021generative} can be incorporated to identity the graph structures (node/edge/subgraph) that trigger \model to conclude that $B$ does not satisfy $\phi$, which brings better interpretation of \model. We believe this would further solidify \model's claims and reduce dependency on traditional model checkers.

\bibliographystyle{ACM-Reference-Format}
\bibliography{ref}

\newpage
\appendix
\section{Temporal Operator Notations\label{appx:symb}}
\begin{table}
\centering
\caption{Temporal Operators in Linear Temporal Logic}
\label{tab:symbol}
\centering
\resizebox{0.8\columnwidth}{!}{
  \begin{tabular}{lcccccccc}
    \toprule
    \textbf{Symbol}&G&F&R&W&M&X&U\\
    \midrule
    \textbf{Meaning} & globally & finally & release & weak until & strong release & next & until \\
  \bottomrule 
\end{tabular}}
\vspace{-3mm}
\end{table}

The meaning of temporal operators supported by $\phi$ is presented in Table \ref{tab:symbol}.

\section{Experimental Settings \label{appx:exp}}
\subsection{Environment} 
Experiments were performed on a cluster with four Intel 24-Core Gold 6248R CPUs, 1TB DRAM, and eight NVIDIA QUADRO RTX 6000 (24GB) GPUs. 

\subsection{Dataset Description \label{appx:data}}
The specifications ($\phi$) for \texttt{Short} and \texttt{Diverse} are constructed by Spot. The corresponding systems ($B$) for specifications of both Spot generated and RERS are created by LTL3BA. 

These three datasets are constructed to test the generalization capability of \proj across different distributions and varying length specifications. The distribution of \texttt{Diverse} subsumes \texttt{Short} and \texttt{RERS19}, and comprises of the most complex and lengthy specifications and corresponding systems, for which LTL3BA takes the longest time. Dataset \texttt{Diverse} is designed to test the performance of \model on diversified samples, where the ability of generalization is examined via varying length systems and specifications presented in the evaluation across different scenarios, where previous related works such as \cite{ZhuMC} are suffered from.  

\subsubsection{Spot Generated Datasets} 
\begin{figure*}
\subfigure[Length distribution \label{fig:lenShort}]{
\begin{minipage}{0.45\textwidth}
\centering    
\includegraphics[height=4cm]{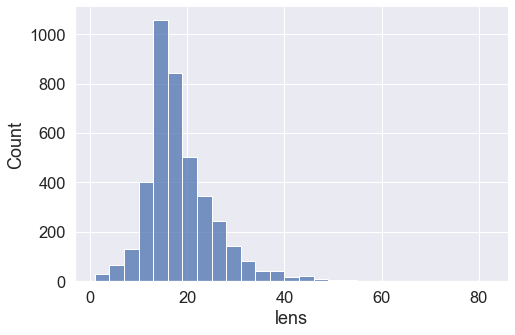}
\end{minipage}
}
\subfigure[Length v.s. State distribution \label{fig:stateShort}]{
\begin{minipage}{0.45\textwidth}
\centering
\includegraphics[height=4.5cm]{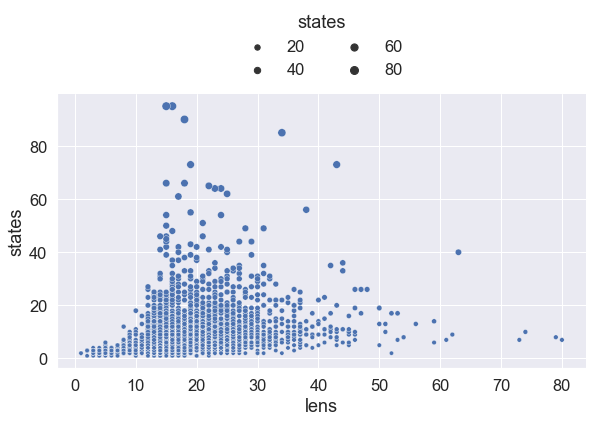}
\end{minipage}
}

\subfigure[Length v.s. Edge distribution \label{fig:edgeShort}]{
\begin{minipage}{0.45\textwidth}
\centering    
\includegraphics[height=4cm]{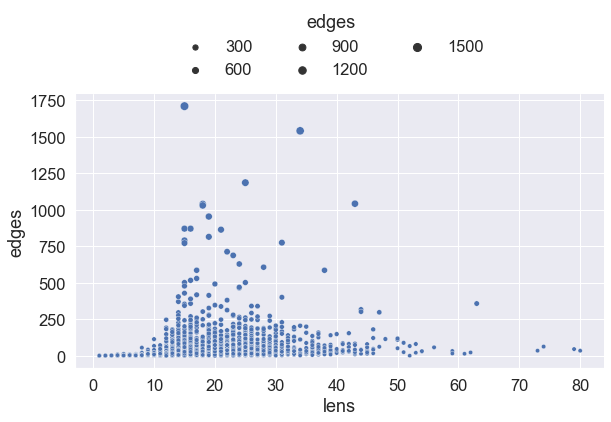}
\end{minipage}
}
\subfigure[Edge v.s. State distribution \label{fig:stateEdgeShort}]{
\begin{minipage}{0.45\textwidth}
\centering
\includegraphics[height=4.5cm]{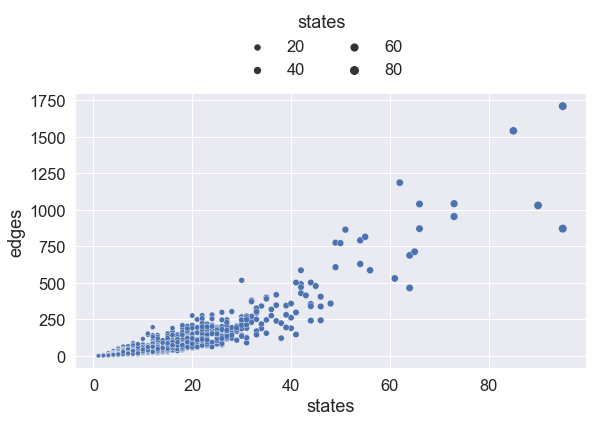}
\end{minipage}
}
\caption{Statistical Summary of Dataset \texttt{Short}.}
\label{fig:scalingShort}
\end{figure*}

\begin{figure*}
\centering
\subfigure[Length distribution \label{fig:lenDiverse}]{
\begin{minipage}{0.45\textwidth}
\centering    
\includegraphics[height=4.5cm]{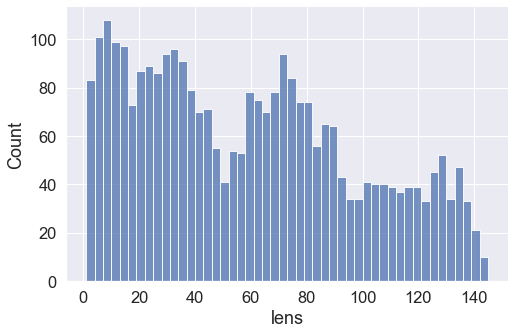}
\end{minipage}
}
\subfigure[Length vs State distribution \label{fig:stateDiverse}]{
\begin{minipage}{0.45\textwidth}
\centering
\includegraphics[height=4.5cm]{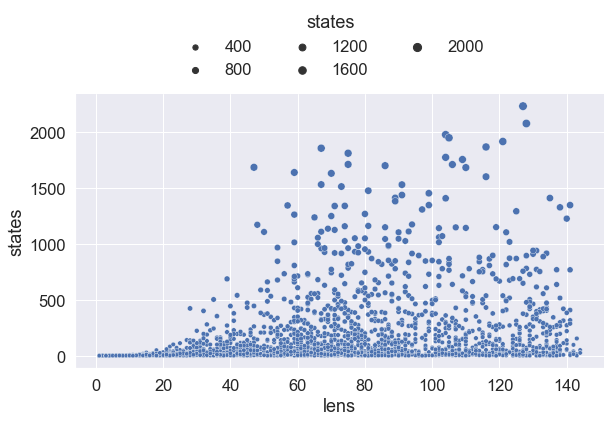}
\end{minipage}
}
\centering
\subfigure[Length vs Edge distribution \label{fig:edgeDiverse}]{
\begin{minipage}{0.45\textwidth}
\centering    
\includegraphics[height=4.5cm]{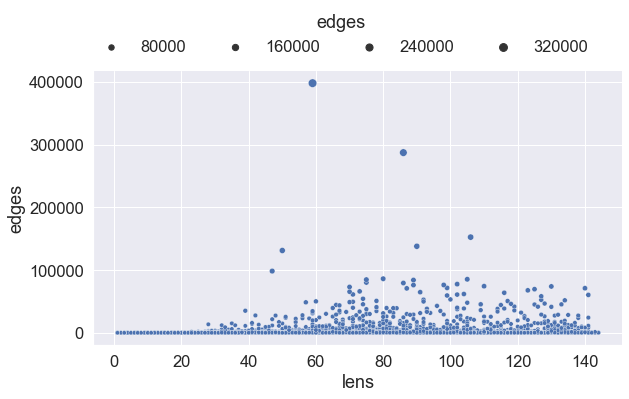}
\end{minipage}
}
\subfigure[Edge vs State distribution \label{fig:stateEdgeDiverse}]{
\begin{minipage}{0.45\textwidth}
\centering
\includegraphics[height=4.5cm]{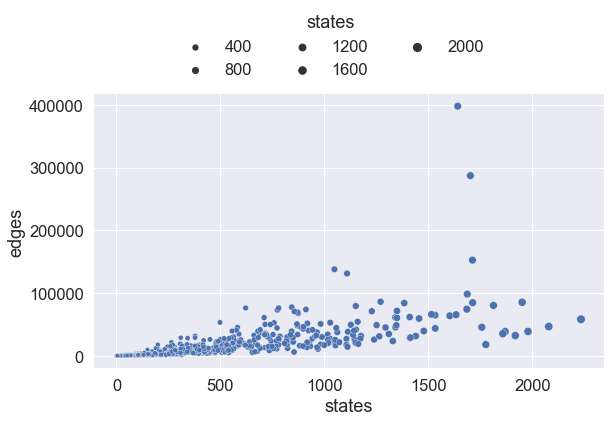}
\end{minipage}
}
\caption{Statistical Summary of Dataset \texttt{Diverse}.}
\label{fig:scalingDiverse}
\end{figure*}

The \texttt{randltl} feature of Spot controls the length of generated LTL formulae, where the default size of expression tree is set to 15. The output formulae are not syntactically the same and less than 10\% of them are semantically equivalent. There are two types of datasets generated through Spot: \texttt{Short}, where the length of the LTL formulae (noted as \#Lens) range from 1 to 80 and \texttt{Diverse}, where the length of the LTL formulae range from 1 to 144. The length distribution of the formula for both \texttt{Short} and \texttt{Diverse} is plotted in Figures \ref{fig:lenShort} and \ref{fig:lenDiverse}. The number of states and edges of these two datasets are described in Figures \ref{fig:stateShort}, \ref{fig:edgeShort}, \ref{fig:stateDiverse}, and \ref{fig:edgeDiverse}, respectively. By observing those distributions, it can be concluded that \texttt{Diverse} comprises of the largest automata and an uniform length distribution of LTL formulae. The range of length and states of LTL formulae is similar between \texttt{Short} and \texttt{RERS19}, but the corresponding transition range is less than 160 for \texttt{RERS19} while 1,711 for \texttt{Short}. The distribution pattern of both $\phi$ and $B$ for dataset \texttt{Diverse} is quite different from the other two datasets.

\begin{figure*}
\subfigure[Length distribution \label{fig:lenRERS}]{
\begin{minipage}{0.45\textwidth}
\centering
\includegraphics[height=4.5cm]{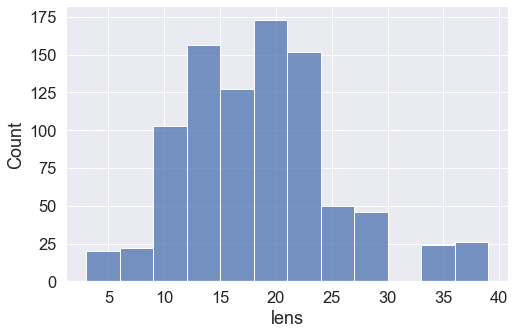}
\end{minipage}
}
\hfill
\subfigure[Length vs State distribution \label{fig:stateRERS}]{
\begin{minipage}{0.45\textwidth}
\centering
\includegraphics[height=4.5cm]{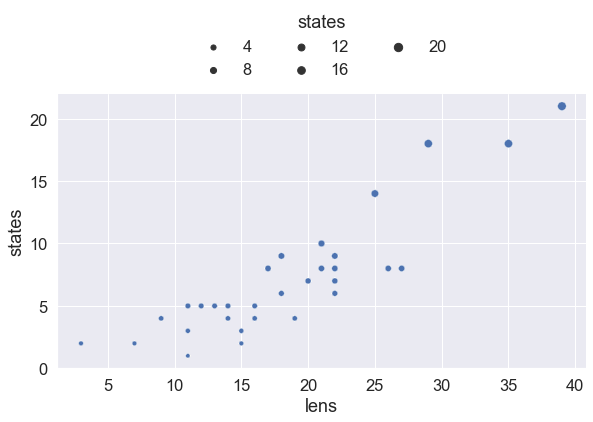}
\end{minipage}
}
\hfill
\subfigure[Length vs Edge distribution \label{fig:edgeRERS}]{
\begin{minipage}{0.45\textwidth}
\centering    
\includegraphics[height=4.5cm]{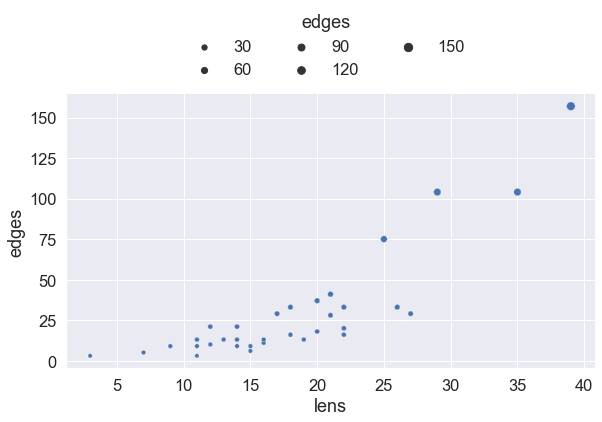}
\end{minipage}
}
\hfill
\subfigure[Edge vs State distribution \label{fig:stateEdgeRERS}]{
\begin{minipage}{0.45\textwidth}
\centering
\includegraphics[height=4.5cm]{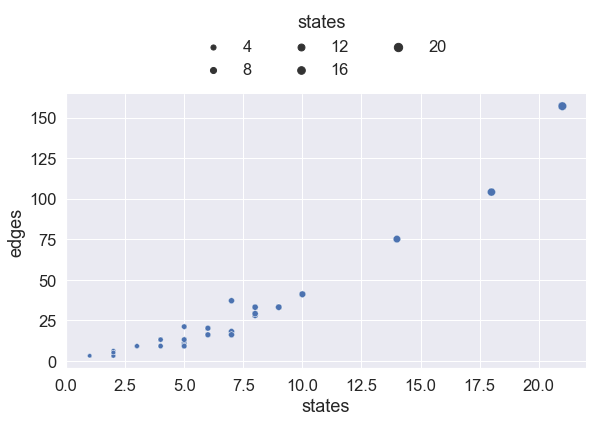}
\end{minipage}
}
\caption{Statistical Summary of Dataset \texttt{RERS19}.}
\label{fig:scalingRERS}
\end{figure*}

\subsubsection{RERS Dataset}
\textbf{Rigorous Examination of Reactive Systems (RERS)} is an international model checking competition track organized every year. We adopt 1800 problems from the Sequential LTL track and compare the performance of \proj with respect to the top 3 teams on the leaderboard, in terms of accuracy under time constraints. The statistical details of dataset \texttt{RERS19} is presented in Figure \ref{fig:scalingRERS}.

\newpage
\subsection{Architecture and Hyperparameters}
\begin{figure}[htp]
    \centering
    \includegraphics[width=\textwidth]{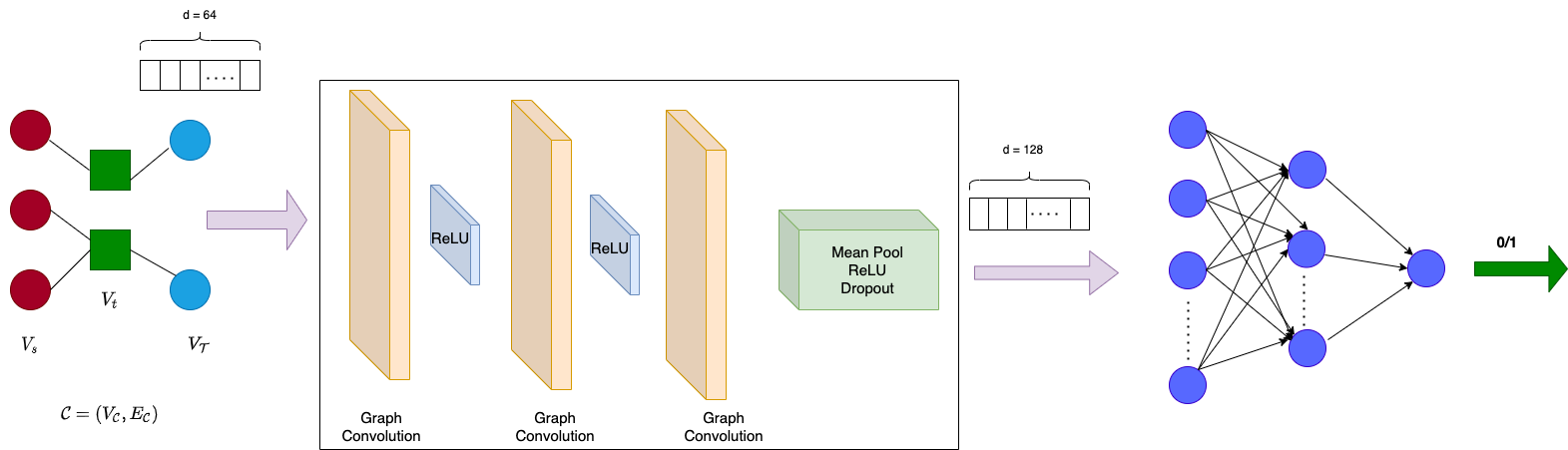}
    \caption{Overview of \proj Architecture.}
    \label{fig:OCTAL}
\end{figure}

The architecture of \proj (illustrated in Figure \ref{fig:OCTAL}) comprises of a three-layer GNN, followed by a two-layer MLP to classify whether $B$ satisfies $\phi$. Mean pooling is used to aggregate the learnt node embeddings. A dropout rate of 0.1 is used, along with 1D batch normalization in every convolution layer of GNN. ReLU \cite{RelU18} is used as the non-linear activation between GNN and MLP layers. Every node in $\mathcal{C}$ has an initial embedding of length 64 (or 66 for the directed $\mathcal{G}$). The GNN framework produces an embedding for $\mathcal{C}$ of dimension 128 after mean pooling. MLPs take this hidden graph representation as the input, and produce a `0/1' result as the final prediction.

\subsection{Implementation Details}
The code base is implemented on PyTorch 1.8.0 and pytorch-geometric \cite{pytorchGeometric}. The source files are attached along with the submitted supplementary.

\section{Additional Experimental Results \label{appx:res}}

\begin{figure*}
\subfigure[Length \label{fig:dlen}]{
\begin{minipage}{0.31\textwidth}
\centering
\includegraphics[width=4cm]{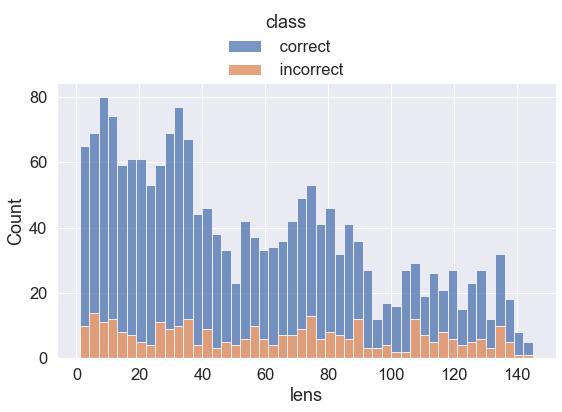}
\end{minipage}
}
\hfill
\subfigure[States \label{fig:dstate}]{
\begin{minipage}{0.31\textwidth}
\centering
\includegraphics[width=4.5cm]{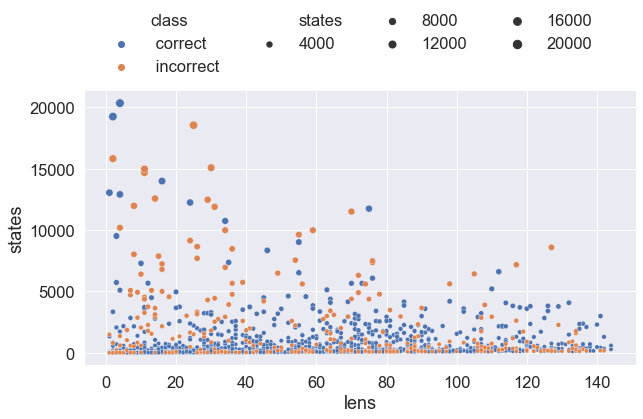}
\end{minipage}
}
\hfill
\subfigure[Edges \label{fig:dEdge}]{
\begin{minipage}{0.31\textwidth}
\centering
\includegraphics[width=4.5cm]{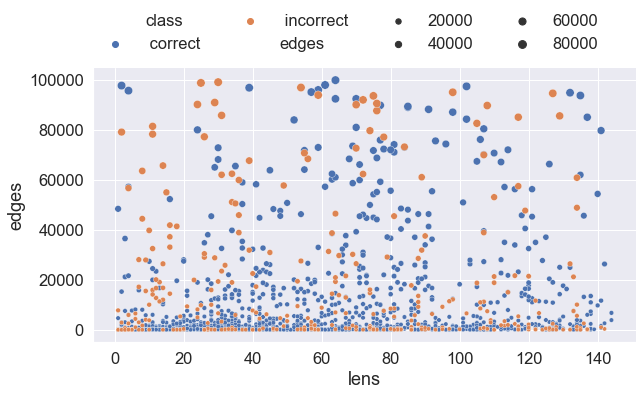}
\end{minipage}
}
\vspace{-3mm}
\caption{Distribution of Predicted Results on Dataset \texttt{Diverse}. \label{fig:diversePrediction}}
\end{figure*}

\subsection{Generalization on Dataset \texttt{Diverse}}
We further tested the generalization of \proj on model checking problems with a larger and diverse distribution by training \proj on \texttt{Short} and testing with samples in \texttt{Diverse}. \proj(GIN) achieves 84.2\% accuracy on average which is slightly decayed from the performance in Table \ref{tab:acc} but still performs reasonably well across larger $B$'s and longer $\phi$'s. Figure \ref{fig:diversePrediction} shows the distribution of correctly and incorrectly predicted samples. \proj generalizes pretty well across varying length $\phi$ and for $B$'s of sizes much larger than the range of \texttt{Short}. The distribution pattern for the correct prediction is similar to \texttt{Diverse}'s as shown in Figure \ref{fig:scalingDiverse}, which indicates \proj can well generalize on unseen larger samples in testing. 

\begin{table}[t]
\caption{Classification accuracy and precision/recall for one-hot encoding. \texttt{RERS19(S)} and \texttt{RERS19(D)} represent models being trained on \texttt{Short} and \texttt{Diverse}, and then tested on \texttt{RERS19}.}
\label{tab:accOneHot}
\label{tab:precrecallDirected}
\centering
\renewcommand{\arraystretch}{1.15} 
\resizebox{0.75\columnwidth}{!}{
\begin{tabular}{ll|cc|cc}
\toprule
    & \textbf{Models} & \texttt{Short} & \texttt{Diverse} & \texttt{RERS19(S)} & \texttt{RERS19(D)} \\
\midrule
    & \textbf{OCTAL} & 95.60±0.31 & 86.74±0.71 & 94.95±0.71 & 94.90±0.56\\
\midrule
\midrule
    \multirow{3}{*}{\rotatebox{90}{~~~~\texttt{OCTAL}}} & \textbf{Precision} &  95.47±0.39 & 86.26±1.54 & 97.02±0.59 & 92.55±0.90\\
    & \textbf{Recall} &  95.75±0.77 & 86.45±1.31 & 92.75±1.38 & 97.66±0.91\\
\bottomrule
\end{tabular}}
\vspace{-2mm}
\end{table}

\begin{table}[t]
\caption{Classification accuracy and precision/recall for Directed $G$. \texttt{RERS19(S)} and \texttt{RERS19(D)} represent models being trained on \texttt{Short} and \texttt{Diverse}, and then tested on \texttt{RERS19}.}
\label{tab:accDirected}
\vspace{0mm}
\centering
\renewcommand{\arraystretch}{1.15} 
\resizebox{0.75\columnwidth}{!}{
\begin{tabular}{ll|cc|cc}
\toprule
    & \textbf{Models} & \texttt{Short} & \texttt{Diverse} & \texttt{RERS19(S)} & \texttt{RERS19(D)} \\
\midrule
    & \textbf{OCTAL} & 95.75±0.77 & 87.55±0.40 & 95.17±1.50 & 94.48±0.51\\
\midrule
\midrule
    \multirow{3}{*}{\rotatebox{90}{~~~~\texttt{OCTAL}}} & \textbf{Precision} &  95.13±1.20 & 85.69±1.51 & 97.90±1.15 & 92.49±1.59\\
    & \textbf{Recall} &  97.20±0.55 & 89.06±2.50 & 92.37±4.08 & 96.88±1.89\\
\bottomrule
\end{tabular}}
\vspace{-2mm}
\end{table}

\subsection{Settings for One-hot Encoding}
Instead of sampling from a normal distribution as described in Section \ref{sec:encoding}, each variable/operator is represented with the value `1' in its respective position, if $v$ contains the representation for it. Otherwise, the position has the value `0'. The rest rules for encoding remain the same.

\subsection{Settings for \texorpdfstring{$\mathcal{G}$} Directed}
In addition to the node encoding presented in Section \ref{sec:encoding}. Part VI is added and encodes the direction of state transitions for directed $\mathcal{G}$ as following,

\[
      \{\substack{\RN{1} \\ \underbrace{[\_]}_{1/0}}
      \substack{\RN{2} \\ \underbrace{[\_,\_,\_,....,\_]}_{\forall a \in \mathcal{A}}}
      \substack{\RN{3} \\ \underbrace{[\_,\_,\_,....,\_]}_{!a/\neg a, \forall a \in \mathcal{A}}}
      \substack{\RN{4} \\ \underbrace{[\_,\_,\_,....,\_]}_{ \forall o \in \{\mathcal{O} / !\}}}
      \substack{\RN{5} \\ \underbrace{[\_,\_]}_{q \in Q} \}}
      \substack{\RN{6} \\ \underbrace{[\_,\_]}_{v \in V_{s}}} \}
\]
In this case, the direction needs to be encoded only in $\mathcal{G}$, which is the bipartite representation of $B$. Every $v \in V_{e}$, stores the source and destination vertices of the corresponding transition. The first field represents the source $v_i \in V_{s}$, and the second filed records the destination $v_j \in V_{s}$. Part VI of $V_{s}$ and $V_{\mathcal{T}}$ would be all zeros as the direction of transitions is sufficiently encoded in $V_{e}$.

\subsection{Analysis of Additional Results}
The experimental results of one-hot encoding and Directed $G$ are presented in Tables \ref{tab:accOneHot} and \ref{tab:accDirected}, respectively. The experiments were run under similar settings for five independent runs, with mean and standard deviation reported for accuracy, precision and recall. Observed from those tables, \model performs roughly well in general on all configurations of both one-hot and directed $\mathcal{G}$. For one-hot encoding, it only slightly outperforms the original encoding on domain shift for \texttt{RERS19} but could not obtain further gains on other two datasets as shown in Table \ref{tab:accOneHot}. On comparison with the undirected $\mathcal{G}$ presented in Table \ref{tab:acc}, \model performs slightly better in general on \texttt{Diverse} with the directed $\mathcal{G}$. The rest of the configurations give comparable performance in terms of accuracy, precision and recall with respect to Table \ref{tab:accDirected}. From the results reported above, we can conclude that the current position specific representation for variables, their negated form, and operators are sufficient for \proj to obtain consistent performance, along with robust enough generalization capabilities and balanced computational cost.

\end{document}